\documentclass{ws-procs975x65}

\begin{document}

\title{Black Holes From the Dark Ages: \\
Exploring the Reionization Era \\
and Early Structure Formation \\
With Quasars and Gamma-Ray Bursts}

\author{S. G. Djorgovski}

\address{Division of Physics, Mathematics, and Astronomy, 
Caltech, Pasadena, CA 91125, USA.
email: george@astro.caltech.edu}

\maketitle

\abstracts{
The cosmic reionization era, which includes formation of the first
stars, galaxies, and AGN, is now one of the most active frontiers of 
cosmological research.  We review briefly our current understanding
of the early structure formation, and use the ideas about a joint
formation of massive black holes (which power the early QSOs) and
their host galaxies to employ high-redshift QSOs as probes of the
early galaxy formation and primordial large-scale structure.  There
is a growing evidence for a strong biasing in the formation of
the first luminous sources, which would lead to a clumpy reionization.
Absorption spectroscopy of QSOs at $z \ge 6$ indicates the end
of the reionization era at $z \sim 6$; yet measurements from the
WMAP satellite suggest and early reionization at $z \sim 10 - 20$.
The first generation of massive stars, perhaps aided by the early
mini-quasars, may have reionized the universe at such high redshifts,
but their feedback may have disrupted the subsequent star and galaxy
formation, leading to an extended and perhaps multimodal reionization
history ending by $z \sim 6$.  Observations of $\gamma$-ray bursts
from the death events of these putative Population III stars may 
provide essential insight into the primordial structure formation,
reionization, early chemical enrichment, and formation of seed
black holes which may grow to become central engines of luminous
quasars.
}

\section{Introduction.  The End of the Cosmic Dark Ages}

Exploration of the cosmic reionization era -- from the appearance of
the first luminous sources to the end of the phase transition when
the intergalactic hydrogen becomes fully reionized and transparent
to the UV light -- is now rapidly becoming an active frontier of 
cosmological research.
\footnote{We use the term reionization {\it era} for an extended period
of time lasting a few hundred million years, from the formation of the
first luminous sources at $z \sim 20-30$, to the final conversion of
the cosmic hydrogen from neutral to ionized state at $z \sim 6$, 
rather than a much shorter period of time when that phase transition
is completed universe-wide (a somewhat ill-defined concept), which we
could call the reionization {\it epoch}.} 
This fundamental cosmological era signals
the appearance of first luminous sources, including the first population
of stars and their explosive ending events, the first galaxies, and
the early AGN.  In this review we describe some the current ideas
and recent results in this field, with an emphasis on the use of
high-redshift quasars (QSOs) and $\gamma$-ray bursts (GRBs) as
probes of the early structure formation and reionization.

There are two main streams of cosmology: the traditional one, which
originated with Hubble, which aims to measure the values of parameters
which describe the global geometry and kinematics of the universe;
and a quest to understand the formation and evolution of the major
constituents of the universe on large scales, galaxies and the large-scale
structure (LSS).  The task of the former is now essentially complete,
with the values of cosmological parameters measured with a percent-level
precision from the studies of the cosmic microwave background (CMBR)
fluctuations, along with the data from SN Hubble diagrams, LSS
surveys, etc.  There is also an elegant tributary stream of physical
cosmology which tackles the early universe, from the Planck
era, through the inflation era, cosmic nucleosynthesis, and 
a detailed analysis of the CMBR fluctuations at the recombination
epoch.  But the second major stream of cosmology, understanding of
the structure formation and evolution, remains as a messy, difficult,
and challenging enterprise.

Nevertheless, there has been a remarkable progress in our understanding
of galaxy and structure formation in the past several years.  The general
hierarchical picture of galaxy and structure formation is now firmly
established, both observationally and theoretically.  However, as is always
the case in cosmology, the clarity and certainty of our knowledge fade as
the look-back time and distance increase.  While we have a fairly solid
grasp of galaxy evolution out to $z \sim 1$, and some insights out to
$z \sim 4 - 5$, the early, formative stages of the cosmic structure, say
within the first 1 Gyr after the Big Bang, remain murky and are now becoming
a focal arena of cosmological research.

An increasing number of necessarily young (due to their high redshifts)
galaxies are being discovered.  As of mid-2004, there are
several tens of spectroscopically confirmed objects at $z > 5$
(e.g., Dey et al.~1998; Weymann et al.~1998; Spinrad et al.~1998;
Ellis et al.~2001; Dawson et al.~2002; Rhoads et al.~2003; Stanway et al.~2004;
Hu et al.~1998, 2004; etc.),
and a rapidly growing number of objects at $z > 6$
(e.g., Hu et al.~2002; Cuby et al.~2003; Kodaira et al.~2003; Kneib et al.~2004;
Kurk et al.~2004; Stern et al.~2004; etc.), with an even larger number
of color-selected candidates which may be at comparable redshifts.
These discoveries are providing us with some direct
insights into the early stages of galaxy and LSS formation.

\begin{figure}[ht]
\centerline{\epsfxsize=4.9in\epsfbox{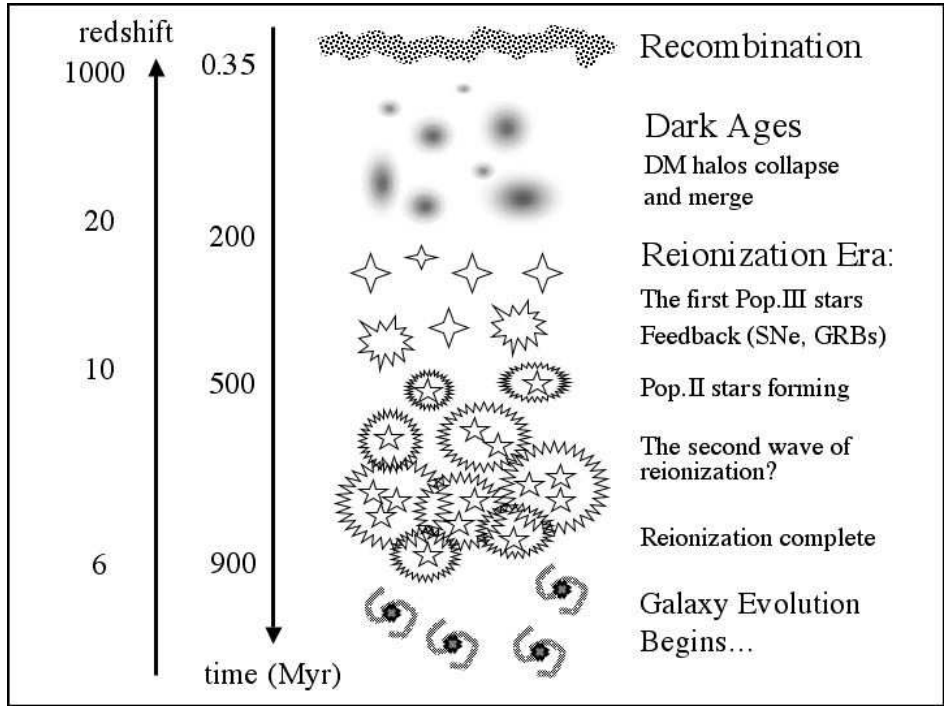}}   
\caption{
A schematic outline of the early cosmic history, according to our current
understanding.  Approximate redshifts of key epochs are indicated on the
left, and converted to the corresponding ages using the now standard
Friedmann-Lemaitre cosmology with $h = 0.7$, $\Omega_0 = 0.3$, and
$\Lambda_0 = 0.7$.  Needless to say, we have a very poor understanding of
what really happened during the reionization era, and when it actually
started; but we are pretty sure that it ended by about $z \sim 6$.
\label{} }
\end{figure}

Theoretical reviews include, e.g., Loeb \& Barkana (2001), Barkana \& Loeb 
(2001), and Miralda-Escud\'e (2003a).
In a nutshell, the currently believed scenario is as follows.
When the Universe was about 380,000 years old,
at a redshift $z_{rec} \sim 1100$, it underwent a phase transition,
from an incandescent plasma containing the heat of the Big Bang, to a space
filled with dark matter, energy and neutral gas.  The Universe then entered the
``dark ages'', with embryonic structures growing from the seeds of dark-matter
fluctuations, which are observed today as ripples in the CMBR. After a few
hundred million years, these condensations became dense enough for the first
stars to form, leading to the appearance of the first galaxies and the growth of
the massive black holes that are believed to power quasars.

As these first objects lit up, they also modified the gas between them, ionizing
the hydrogen and making it transparent to ultraviolet light.  The
universe underwent another phase transition, from a neutral to an ionized state.
Each of the primordial sources of light ‹- most powered by young, massive stars,
but some powered by the accretion of matter into growing black holes (the early
quasars) ‹- excavated a Str\"omgren bubble of ionized gas in the otherwise
neutral surrounding medium.  When these bubbles began to percolate,
the reionization was complete.  

The standard observational test of the approach to reionization is the
prediction of an extended, optically thick absorption due to neutral hydrogen
at $\lambda_{rest} < 1216$ \AA\ (Gunn \& Peterson 1965; hereafter GP), as
distinct from the usual Ly$\alpha$ forest due to a multitude of distinct
hydrogen clouds embedded in a largely ionized intergalactic medium (IGM)
as observed at lower redshifts.  Indeed, the GP effect was recently observed
in a spectrum of quasar SDSS 1030+0524 at $z = 6.28$ (Becker et al.~2001; Fan
et al.~2001; see also Pentericci et al.~2002, Djorgovski et al.~2001, 2002, and
in prep.).  Such
absorption troughs are now seen along every line of sight to QSOs at $z > 6$.
However, there are some interpretative complications remaining.

The first protogalactic starbursts probably appeared at $z \sim 20 - 30$. 
They may have reionized the universe by $z \sim 15 \pm 5$, as indicated by the
recent WMAP measurements (Kogut et al.~2003).  The history of the IGM, ionizing
flux, and early galaxy and AGN formation from that point on is extremely
uncertain and was probably very complex;
it is even possible that the universe was reionized twice (Cen 2003;
Wyithe \& Loeb 2003b; Sokasian et al.~2004; etc.).
The evolution of the mean ionizing 
flux with redshift is expected to be very strong, due to the gradual 
appearance and growth of the sources (protogalaxies and early AGN) 
responsible for the reionization.

It is now generally believed that the reionization sources
were mainly star-forming galaxies rather than AGN, and recent studies
indicate an already vigorous star formation at $z \sim 6$
(Giavalisco et al.~2004; Dickinson et al.~2004; Stanway et al.~2004;
Bouwens et al.~2004; Bunker et al.~2004; Stiavelli et al.~2004; etc.).
However, it is also possible that a substantial
early AGN (``mini-quasar'') activity may predate the peak epoch of star
formation in galaxies (Silk \& Rees 1998; Madau \& Rees 2001). 

The reionization era is a major cosmological milestone.  It signals the
appearance of the first galaxies and AGN, in the epoch of ``cosmic
renaissance''.  The spatial and temporal stucture of the IGM phase transition
reflects the primordial large-scale structure and the early luminosity and
density evolution of the reionization sources, the primordial starbursts and
AGN.  A good understanding of the structure and extent of the reionization is
important by itself, reflecting the earliest phases of structure formation,
and for modeling of CMBR foregrounds at high angular frequencies,
a subject of an increasing cosmological interest.
We are now just starting to probe this fundamental cosmological era.

\section{The Quasar-Protogalaxy Connection}

There are now several compelling and growing lines of evidence that there
are fundamental connection between the formation and evolution of galaxies and
their central massive black holes, which presumably powered the early quasar
activity.  Along with the established uses of QSOs as absorption probes of the
IGM, this suggest their use as direct probes and markers of sites of early,
massive galaxy formation and primordial LSS.  For a review and
references, see, e.g., Djorgovski (1999).

Physically, it is now believed that the same kind of processes, i.e.,
dissipative mergers and tidal interactions, may be fueling both bursts of star
formation and AGN activity (see, e.g., Norman \& Scoville 1988, Sanders
et al.~1988, Carlberg 1990, Hernquist \& Mihos 1995, Mihos \& Hernquist 1996,
Franceschini et al.~1999, Monaco et al.~2000, Kauffmann \& Haehnelt 2000, etc.;
see, e.g., Haehnelt 2004 for a recent review). 

Quasars thus may be a common phase of the early formation of ellipticals and
massive bulges.  QSO demographics support this idea (Small \& Blandford 1992,
Chokshi \& Turner 1992, Haehnelt \& Rees 1993; see Richstone 2004 for a review).
There should be dead (actually only sleeping) QSO remnants, 
supermasive black holes (SMBH), in most of the
massive galaxies today.

Indeed, most or all ellipticals and massive bulges at $z \sim 0$ seem to contain
central massive dark objects, presumably SMBH,
suggestive of an earlier QSO phase (see, e.g., Kormendy 2004, Kormendy \&
Richstone 1995, for reviews and references). Their masses correlate in a roughly
direct proportion with the masses of luminous old stellar components of their
host galaxies (Magorrian et al.~1998; Ferrarese \& Merritt 2000; Gebhardt et al.
2000; Merritt \& Ferrarese 2001),
with typicall mass fractions $M_\bullet / M_\star \sim 10^{-3}$. 
The inferred mean comoving density of these putative QSO remnants,
$\rho_\bullet \sim 5 \times 10^5 M_\odot {\rm Mpc}^{-3}$, is also in a
good agreement with the completely independent estimates obtained from
integration of the total QSO light out to high redshifts, with reasonable
assumptions about the radiative efficiency of mass accretion
(Soltan 1982, Small \& Blandford 1992, Salucci et al.~1999, etc.).

Moreover, there are even better correlations with the stellar dynamics of the
hosts on a scale of a few kpc, $M_\bullet \sim \sigma_\star ^{4.5}$
(Ferrarese \& Merritt 2000; Gebhardt et al.~2000),
whereas the SMBHs have radii of the order of $10^{-4}$ pc);
and with the masses of the host galaxy halos, on the scales of $\sim 10^5$ pc,
with typical $M_\bullet / M_\star \sim ~{\rm few}~ \times 10^{-5}$
(Ferrarese 2002a).  It is intriguing that the observed slope of this relation,
$M_\bullet \sim M_{halo}^{1.6}$ is close to that predicted in the model by
Haehnelt et al.~(1998).  Note that the relation implies that 
more massive -- and also less dense -- host halos are much more
effective in forming and/or growing SMBHs; and there may be even a threshold
halo mass needed for a presence of a SMBH.  For a good review of these
issues, see, e.g., Ferrarese (2002b).
A successful theoretical model needs to explain both the slopes of these
relations, and account for a remarkably small intrinsic scatter.
In any case, these correlations strongly suggest a co-formation and
co-evolution of galaxies and their central SMBHs, with some as-yet poorly
understood feedback mechanisms playing an important role.

Additional clues come from the measurements of the metallicities and
abundances in QSO broad emission line regions.  High metallicities
(up to $10 \times ~Z_\odot$!) observed in $z > 4$ quasars
(Hamann \& Ferland 1993, 1999; Matteuci \& Padovani 1009; Dietrich et al.~2003a) are indicative of a considerable chemical
evolution involving several generations of massive stars in a system massive
enough to retain and recycle their nucleosynthesis products, e.g., comparable
to the cores of giant ellipticals (Romano et al.~2002). 
This is also supported by detections of copious amounts of molecular gas
and dust in large numbers of high-$z$ QSOs (Guilloteau et al.~1999; Omont et al.
2001; Bertoldi et al.~2003a,b; Walter et al.~2003; etc.)

Intriguingly, the relative abundances of the $\alpha$-group elements, which
are produced predominantly by the Type II SNe (exploding massive stars), and
Fe-group elements, which are contributed mainly by the type Ia SNe (detonating
white dwarfs) show an enhanced Fe/Mg ratios out to $z \sim 5$.  Since there
should be some delay (perhaps a few $\times 10^8$ yr?) between the onset
of star formation and the average explosion time for the progenitors of
Type I SNe, this suggests that the star formation in the hosts of these QSOs
started at even higher redshifts, perhaps at $z > 10$ (Dietrich et al.~2003b),
making them possibly significant contributors to reionization.

Finally, estimates of the SMBH masses powering high-$z$ QSOs indicate that
objects with $M_\bullet > 10^9 M_\odot$ are present even out to $z \sim 5 - 6$
(Willott et al.~2003; Dietrich \& Hamann 2004; McLure \& Dunlop 2004).
This can be alleviated, but only slightly, by invoking beaming (which does
boost the apparent continuum luminosity, but does not affect the line
velocity widths, both of which are used in estimating the $M_\bullet$ in
QSOs), or gravitational lensing (ditto; and we also know that only a small
fraction of bright QSOs, $\sim$ a few $\times 10^{-3}$, are strongly lensed).
While the evolution of the $M_\bullet / M_{halo}$ relation out to such high
redshifts is highly uncertain at this time, these QSOs are likely situated in
very massive hosts, e.g., with $M_{halo} \sim 10^{12} - 10^{13} M_\odot$
(see also Turner 1991 for a prescient discussion of these matters).
Such massive halos should be rare, and may be associated with $\sim 4$ to
$5$-$\sigma$ peaks of the primordial density field, and thus be highly 
biased (Efstathiou \& Rees 1998, Cole \& Kaiser 1989, Nusser \& Silk 1993, etc.),
possibly marking the cores of future rich clusters. 

Overall, there is now a compelling evidence for a co-formation and co-evolution
of SMBHs (QSO engines) and their host galaxies.  More details and references can
be found, e.g., in the proceedings dedicated to this topic, edited by Ho (2004).

\section{The Evidence for a Strongly Biased Early Galaxy Formation}

One essential aspect of the early galaxy and structure formation is that it
should be very uneven spatially, starting at the highest peaks of the primordial
density field, and then spreading out.  This is a generic prediction in
essentially all modern models of galaxy formation, resting only on the
reasonable assumption that star and galaxy formation will start in the densest
regions first.  The most massive density peaks in the early universe, and the
luminous objects forming in or near them, are likely to be strongly clustered
{\it a priori} (Kaiser 1984), and thus represent {\it biased} tracers of the
density field.  {\sl The early formation of galaxies should be closely related
to the primordial large-scale structure.}

\begin{figure}[ht]
\centerline{\epsfxsize=4.9in\epsfbox{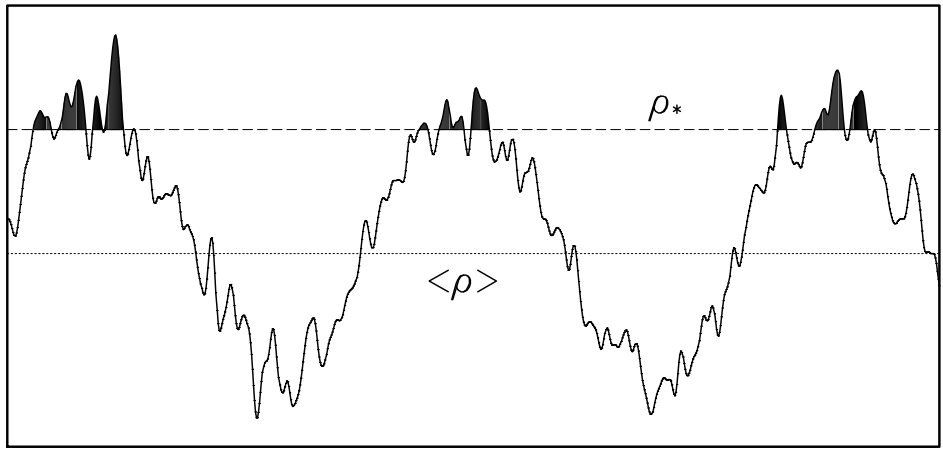}}   
\caption{
A schematic illustration of biasing.  What is shown is a hypothetical cut
through the primordial density field, with a mean value $\langle \rho \rangle$.
Density fluctuations exist on all spatial scales, and the densest spots, say
above some critical density threshold $\rho_*$, 
correspond to the smaller peaks riding atop of the larger waves.  Such dense
spots -- which are the most likely sites of early star, galaxy, and AGN
formation -- are naturally clustered, and may be plausible progenitors of
future rich clusters.  They are the brightest, but highly biased
representatives of the overall density field. As the time goes on, fluctuations
in less dense regions become dense enough to ignite star formation, until the
entire volume is populated by galaxies.
\label{} }
\end{figure}

The evidence for such a bias is already seen with large galaxy samples at 
$z \sim 3 - 3.5$ (Steidel et al.~1998, Adelberger et al.~1998, etc.),
while normal galaxies at $z \sim 0$ seem to be relatively unbiased tracers
of the mass (see, e.g., Lahav et al.~2002, or Verde et al.~2002).
The biasing should be stronger at earlier times, and for the more massive
systems, as indicated in numerous theoretical studies (e.g.,
Brainerd \& Villumsen 1994,  Matarrese et al.~1997, Moscardini et al.~1998,
Blanton et al.~2000, Magliocchetti et al.~2000, Valageas et al.~2001, etc.). 
It is then expected that biasing should be very strong for the
first massive protogalaxies, some of which may be the hosts of quasars at
$z \sim 4 - 6$, and beyond.  Observations indeed confirm this prediction in at
least two ways:

First, an ``excess'' in the number of galaxies and in the density of star
formation (perhaps up to two orders of magnitude more than in the general
field at comparable redshifts) was also seen in a systematic Keck survey of
fields centered on the known $z > 4$ quasars (Djorgovski 1998, 1999; Djorgovski
et al.~1999; and in prep.), consistent with the idea that luminous high-$z$
QSOs may mark sites of future rich clusters of galaxies.

Second, high-$z$ QSOs themselves appear to be strongly clustered. 
The first hints of such an effect at high redshifts were provided by the three
few-Mpc quasar pairs at $z > 3$, found in the statistically complete survey by
Schneider et al.~(1994b), as first pointed out by Djorgovski et al.~(1993), and
subsequently confirmed by more detailed analysis (Kundic 1997, Stephens et al.
1997).  Several serendipitously discovered QSO pairs at $z > 4$ have been found
(Schneider et al.~1994a, Schneider et al.~2000, Djorgovski et al.~in prep.),
and a few-Mpc QSO pair at $z \approx 5$ was found by Djorgovski et al.~(2003a).
Chance probabilities of finding {\it any one} such QSO pair are estimated to
be in the range $\sim 10^{-8} - 10^{-4}$; finding so many of them appears to
be extremely unlikely if QSOs are not physically clustered.
Intriguingly, the frequency of the few-Mpc separation quasar pairs at lower
redshifts is roughly what may be expected from normal galaxy clustering
(Djorgovski 1991; see also Zhdanov \& Surdej 2001).

A substantial decrease in the clustering strength is expected at higher
redshifts, since in any model gravitational clustering is always expected to
grow in time.  Yet, the inferred likely values of the clustering length for
high-$z$ QSOs implied by these observations are in tens of Mpc, much higher
that at $z \sim 0$.  The most natural explanation for this apparent paradox is
that high-$z$ QSOs represent highly biased peaks of the density field, and that
the bias itself evolves in time, as expected from theory.  See, 
Djorgovski et al.~(2003a) for a discussion and references.

The evidence for an increased bias at higher redshifts is not confined to QSOs:
clustering of Ly$\alpha$ emitters at $z \sim 5$ in the Subary Deep Field
(Ouchi et al.~2003, 2004; Shimasaku et al.~2003, 2004).

On even larger physical scales, a very intriguing result was found in the
DPOSS high-$z$ QSO survey (Djorgovski 1998, 1999, and in prep.).
We found an unexpectedly large frequency of QSO pairs and triplets with
the typical comoving separations of $\sim 100 - 200 h^{-1}$ Mpc.  The typical
projected separations of the PSS QSO pairs are $< 2^\circ$, while the
mean surface density at the depth of our survey implies r.m.s. $\sim 10^\circ$.
The statistical significance of this effect is still difficult to quantify,
mainly due to the possible systematics which could, at least in principle,
produce a spurious clustering signal, e.g., slight variations in the survey
depth, completeness of the QSO selection, etc., but it is estimated to be in the
range $\sim 3 - 7 ~\sigma$.

If this is a real superclustering signal, its implications would be profound.
It would represent a primordial (very) large scale structure, delineated by
some of its highest, biased peaks containing quasars, only a few hundred
physical Mpc away from the CMBR photosphere.  It is very intriguing that the
characterictic scale ($\sim 100 - 200 h^{-1}$ comoving Mpc) implied by our
measurements so far is so close to the physical scale of the horizon at the
recombination (i.e., the first Doppler peak in CMBR fluctuations), and the
characteristic scale of superclustering found in at least some redshift surveys
(Broadhurst et al.~1990, Landy et al.~1996).

If the first luminous sources (galaxies and AGN) were strongly clustered due to
biasing, then the structure of the IGM phase transition corresponding to the
reionization was also very clumpy; we examine some evidence for the uneven
reionization below.  This is generally expected from theory (see, e.g.,
Barkana \& Loeb 2004), but we also note that many current numerical simulations
may not be sampling a sufficient comoving volume to probe such effects
(Ciardi et al.~2000, Gnedin 2000, etc.).

\section{Reionization, the Cosmic Renaissance}

The appearance of first luminous sources at $z \sim 20 - 30$ ends the cosmic
dark ages which start at the recombination, at $z \sim 1100$.
As the numbers and luminosities of the first sources increase, ultimately there
are enough ionizing photons produced to reionize most of the intergalactic
hydrogen, a process we now know is complete by about $z \sim 6$ (the
reionization of the intergalactic helium takes a while longer, until the buildup
of the QSO population produces enough harder photons needed for that conversion;
see, e.g., Wyithe \& Loeb 2003a).

The early phases of formation of the first stars, protogalaxies, QSOs, and LSS,
and the early physical and chemical evolution of the IGM are thus intrinsically
connected in a complex interplay of astrophysical processes and feedbacks.
The onset and the finale of the reionization of hydrogen provide
useful and physically meaningful time pegs, and thus we call this cosmic
renaissance the reionization era; but we could just as well call it the
early structure formation era, or the protogalactic era, or something else
equally appropriate.  This era is now becoming the new frontier of observational
and theoretical cosmology.  

As already noted, some combination of the young, massive stars and accretion
processes associated with the first generation of embryonic AGN produces the
necessary ionizing flux for the reionization (e.g., Miralda-Escud\'e 2003b),
with stars almost certainly dominant near the end, at $z \sim 6$.  Direct
searches for ionizing sources at $z \sim 6 - 30$ will thus provide essential
insights into the nature and evolution of this population.

One important question is whether Ly$\alpha$ line can be detected from
objects embedded in a still largely neutral IGM.  Since Ly$\alpha$ photons
injected into a neutral IGM are strongly scattered, the red damping
wing of the GP trough should strongly suppress, or even completely
eliminate, the Ly$\alpha$ emission line (Miralda-Escud\'e 1998;
Miralda-Escud\'e \& Rees 1998; Loeb \& Rybicki 1999). 
Subsequent calculations (e.g., Haiman 2002, Santos 2004)
show that even for faint sources with little
ionizing continuum, a sufficiently broad ($\Delta v \ge 300 {\rm
km}\ {\rm s}^{-1}$) emission line can still remain observable.  The
transmitted fraction depends upon the size of the local cosmological
H II region surrounding a source, and therefore on the ionizing
luminosity and age of the source.  Presence of clustered fainter sources
and possiby also previously active AGN in the vicinity of Ly$\alpha$ emitters
would also facilitate the escape of Ly$\alpha$ photons (Wyithe \& Loeb 2004b). 
Such young galaxies would then be detectable well into the reionization era.

The newly generated population of free electrons also leaves an imprint on
the CMBR through Thompson scattering, and this signal has been indeed observed
by the WMAP satellite (Kogut et al.~2003), suggesting a surprisingly high
optical depth ($\tau \approx 0.17$) and consequently a high redshift for the
reionization epoch, $z \sim 10 - 20$.  It would be interesting to see 
whether the future CMBR observations and analysis confirm this result.

\begin{figure}[ht]
\centerline{\epsfxsize=4.9in\epsfbox{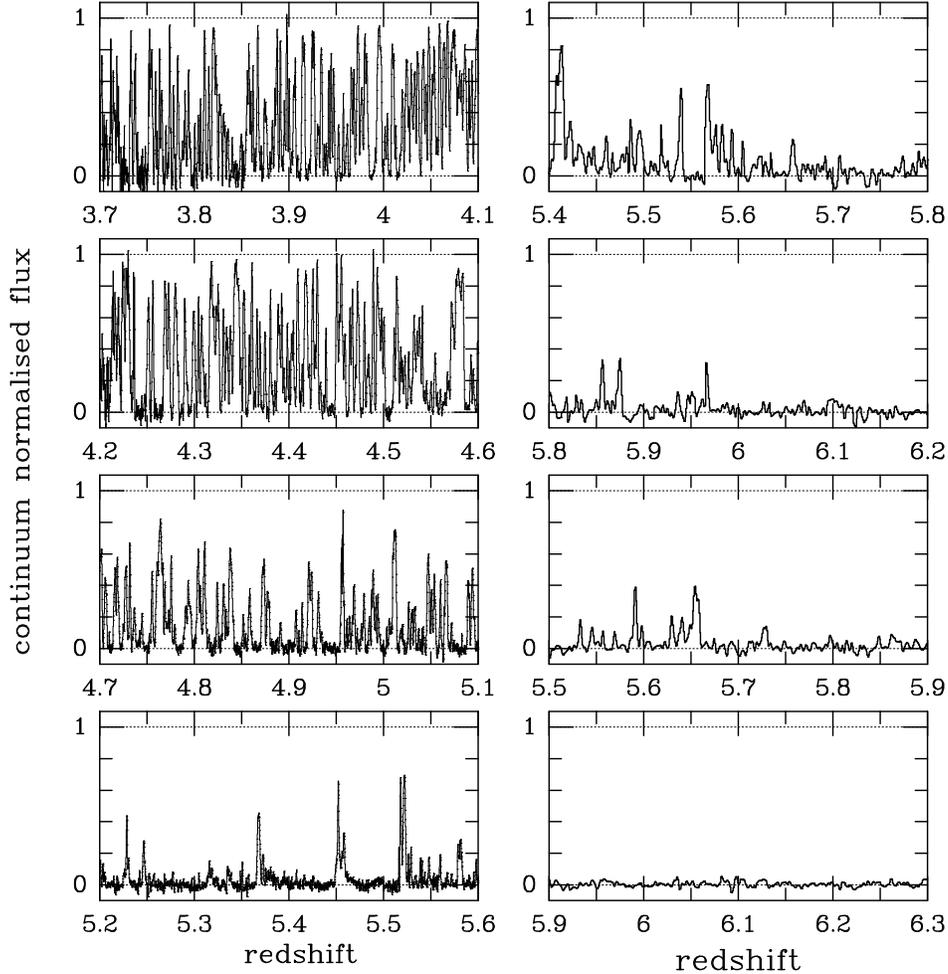}}   
\caption{
Entering the primeval forest: continuum-normalized QSO absorption spectra at
progressively higher redshift windows.  The forest thickens dramatically at 
$z \sim 5.5$ and GP-type troughs are ubiquitous at $z \ge 6$.
Spectra of several high-$z$
SDSS QSOs, obtained at the Keck-II telescope with the ESI instrument were used.
(From Djorgovski et al.~2002, and in prep.)
\label{} }
\end{figure}

QSO absorption studies of the primordial IGM are a complementary probe of the
reionization's end.  As we approach the reionization era from the lower
redshifts, the Ly$\alpha$ forest thickens, with an occasional transmission gap
due to the intersection of ionized bubbles along the line of sight
(Haiman \& Loeb 1999; Loeb 1999); eventually
a complete GP trough is reached.  The signatures of the reionization's finale
have been detected by Becker et al.~(2001), Djorgovski et al.~(2001), and Fan
et al.~(2001, 2003), in the form of extended opaque regions in the spectra of
$z \ge 6$ QSOs. 
However, even small amounts (fraction $\sim 10^{-3}$) of the residual H I can
produce the absorption troughs as those observed so far.  Additional
observational constraints can be obtained, e.g., from the Ly$\beta$ GP effect
(Lidz et al.~2002), and the statistics of absorption window lengths as an
$f(z)$ (Barkana 2002).

\begin{figure}[ht]
\centerline{\epsfxsize=4.9in\epsfbox{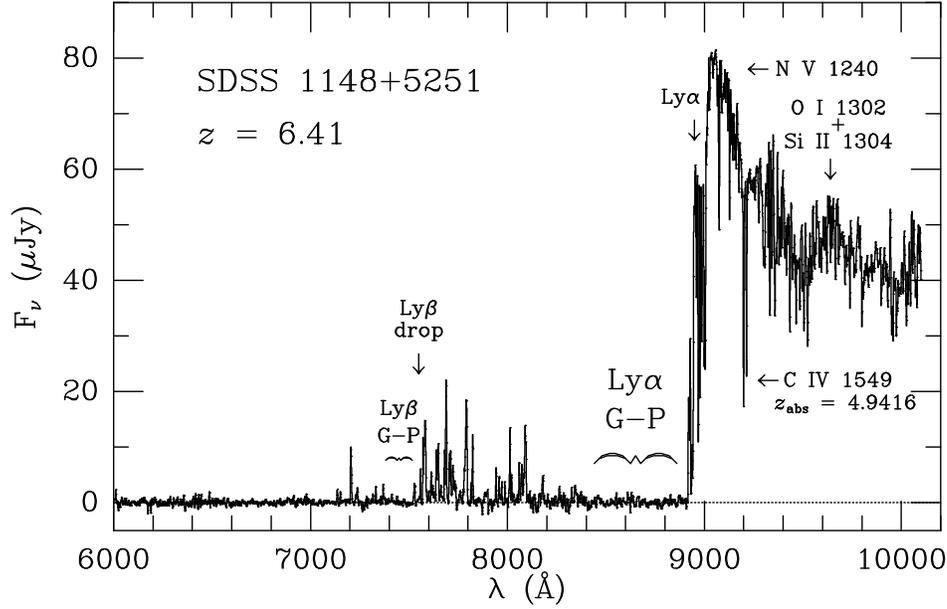}}   
\caption{
The spectrum of SDSS 1148+5251, the currently most distant QSO known, at
$z = 6.41$, obtained at the Keck-II telescope with the ESI instrument.
The GP-type trough is evident blueward of the QSO's Ly$\alpha$ line,
as is its Ly$\beta$ counterpart (both indicated by arrows).  Note also the
carbon absorption doublet from a superposed galaxy at $z_{abs} = 4.9416$;
its continuum emission contaminates the GP signal in this QSO spectrum,
making the measurements of the optical depth difficult.
(From Djorgovski et al.~2002, and in prep.)
\label{} }
\end{figure}

Nevertheless, there are clear indications that some qualitative change
in the state of the IGM occurs at $z \sim 6$:  the GP-like troughs are 
seen along $every$ available line of sight to QSOs at $z >  6$; 
there is a change in the slope of the optical depth vs.
redshift, $\tau(z)$ (Fan et al.~2002; Cen \& McDonald 2002; White et al.~2003;
Djorgovski et al.~2002 and in prep.; however, see Songaila \& Cowie 2002 or 
Songaila 2004 for a contrarian view).
Also, sizes of the observed H II regions around at least some $z > 6$ QSOs
indicate that a substantial fraction of the IGM was neutral at this redshift
(Wyithe \& Loeb 2004a; Mesinger \& Haiman 2004).

\begin{figure}[ht]
\centerline{\epsfxsize=4.9in\epsfbox{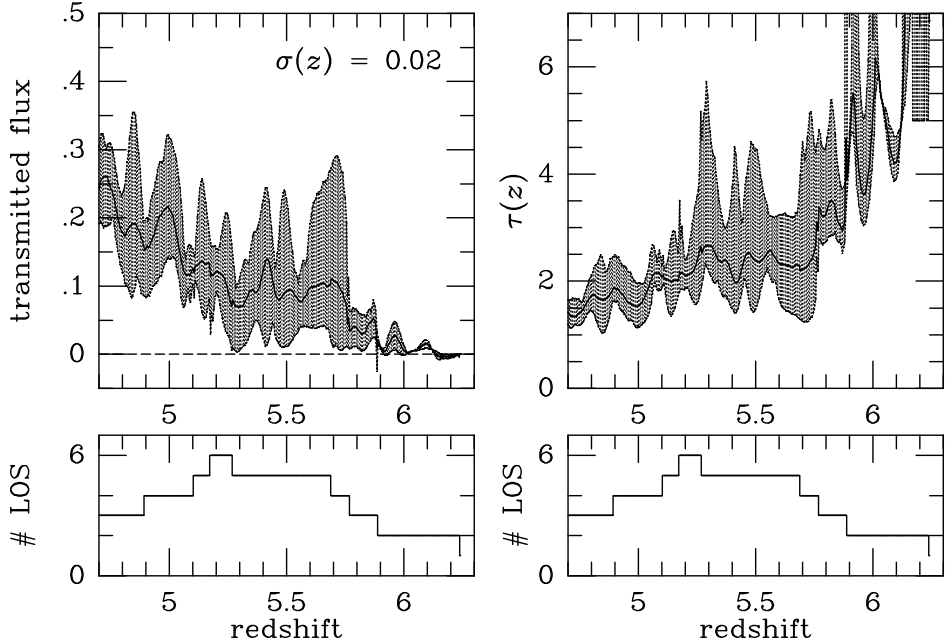}}   
\caption{
The behavior of the transparency of the IGM at high redshifts, shown as the
transmission on the left, and as the optical depth on the right.  The bottom
panels indicate the number of lines of sight used at every redshift.
The solid lines indicate the mean, and the hashed areas the spread among the
different lines of sight.  The data have been smoothed with a Gaussian with
a $\sigma(z) = 0.02$; a wider smoothing or binning, preferred by some authors,
can wash out important features.
Note the change in the slope of $\tau(z)$ around $z \approx 5.8$; this
corresponds to the dramatic change in the ionizing flux and the fraction of
the neutral hydrogen at these redshifts (see, e.g., Fan et al.~2002), which
is interpreted as the signature of the end of the reionization era.
There is also a hint that the spread among different lines of sight (i.e.,
the cosmic variance in the IGM absorption properties) increases just before
the reionization redshift is reached.
(From Djorgovski et al.~2002, and in prep.)
\label{} }
\end{figure}

If {\it both} results are right -- a high-$z$ reionization epoch as indicated
by WMAP, and a $z \approx 6$ reionization epoch indicated by the QSO 
observations, then the history of primordial star, galaxy, and AGN formation
must have been very complex, possibly with multiple waves of reionization, as
suggested, e.g., by Cen (2003), Wyithe \& Loeb (2003b), Haiman \& Holder (2003),
Sokasian et al.~(2004), etc.

Temporal non-uniformity and extended duration of the reionization are a 
natural consequence of the simple fact that the first sources of light did not
all appear at the same time, having some comoving density and luminosity
evolution, with the ionizing flux density gradually increasing in time (see,
e.g., Umemura et al.~2001).  But there might
have been some spatial non-uniformity as well, if the first luminous sources
were a highly biased population, as we argued above.

The inherent non-uniformities (in time and space) of galaxy
and structure formation would be reflected in the structure of the reionization
phase transition.  In addition to the likely strong, bias-driven clustering of
reionization sources (see, e.g., Barkana \& Loeb 2004),
key issues also include the clumpiness of the IGM (e.g., Miralda-Escud\'e et al.
2000), and various feedback mechanisms (see, e.g., Ferrara \& Salvaterra 2004
for a review and references).

\begin{figure}[ht]
\centerline{\epsfxsize=3.5in\epsfbox{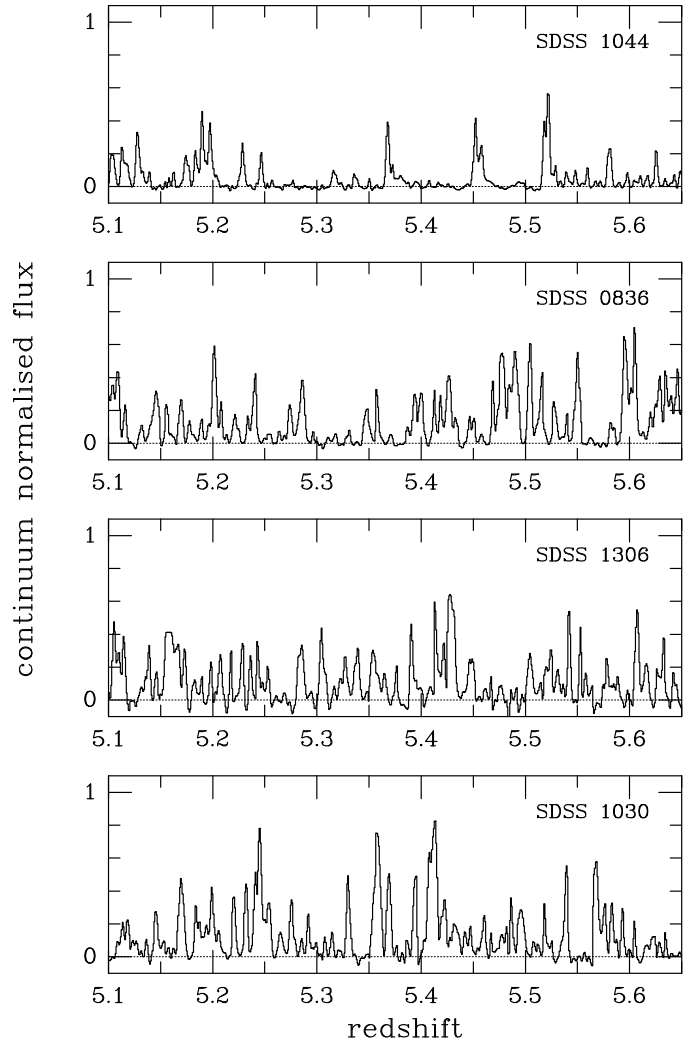}}   
\caption{
A substantial variety in the IGM absorption along 4 different QSO lines of
sight is apparent in these continuum-normalized spectra covering the same
redshift window, which just follows the reionization epoch.  This variance
may be a residual effect of an uneven end to reionization,
as expected from the biased structure formation picture.
(From Djorgovski et al.~2002, and in prep.)
\label{} }
\end{figure}

There is already some evidence for this.  Current
spectroscopic observations of the known QSOs at $z \sim 5.7 - 6.4$ 
(Djorgovski et al.~2002, and in prep.) indicate a
significant variation in the IGM transmission properties along different
lines of sight. 
This is almost certainly a signature of a substantial bias-driven cosmic
variance.  There is a clear need for many more QSO lines of sight,
with high-S/N, high-resolution spectra, in order to probe this effect further.
A combination of the new surveys, such as Palomar-Quest (Djorgovski et al.
2004b), and further work from SDSS will provide a few tens of suitable QSOs
over the next few years.

If there were such large variations in the reionization's
end and aftermath at $z \sim 5 - 6$, then surely there corresponding 
non-uniformities at higher redshifts were even greater, possibly leaving an
imprint on the CMBR fluctuations (see, e.g., Santos et al.~2003). 
This may have some relevance for the
observed excess power above the standard CDM model seen at high angular
frequencies, corresponding to the comoving separations of $\sim 10 - 20$ Mpc
at the CMBR photosphere (Readhead et al.~2004, and references therein).

\section{The First Stars and Gamma-Ray Bursts}

There has been a significant progress in our understanding of primordial star
formation over the past several years.  A generic expectation is that the first,
metal-free (Population III) stars were very massive. 
Unlike the star formation we know about at lower redshifts, the first stars
formed without dust and metals, and probably with much less turbulence and
magnetic fields, unlike their present-day counterparts.  This relative
simplicity gives some hope that their formation can be understood at least
in a broad-brush picture (Bromm, Coppi \& Larson 1999, 2002; Abel, Bryan, \&
Norman 2000, 2002; etc.).
The absence of heavier elements and resulting molecules which now play a major
role in the protostellar cooling, implies that the predominant cooling mechanism
will be through H$_2$ lines.  This, along with other bits of relatively reliable
physics leads to a fragmentation spectrum with a characteristic mass $M_\star
\sim 10^2 - 10^4 M_\odot$ (note that the models do not yet predict an actual
IMF, just a high characteristic mass).  The subject of primordial star
formation has been reviewed recently, e.g., by Bromm (2004).

Such supermassive stars would have typical luminosities
$L \sim 10^6 - 10^7 L_\odot$ and temperatures $T_e \sim 10^5 K$, and be very
effective sources of reionization photons (see, e.g., Bromm, Kudritzki, \& Loeb
2001), possibly reionizing the universe at $z \sim 10 - 20$, in agreement with
the WMAP results (Kogut et al.~2003).  However, their radiative feedback would
effectively prevent formation of any other stars in their vicinity, and their
mechanical feedback may have disrupted their host protogalactic systems,
possibly leading to a substantial slowdown in early star formation, and even
a second recombination from a subsequent Population II stars
(Cen 2003, Wyithe \& Loeb 2003b).  Their ejected metals
would change the physics of the ISM cooling, and the subsequent formation of
Population II stars may have assumed a more normal (i.e., $\sim$ present day)
IMF (Bromm 2004; Yoshida et al.~2004).

It is almost certain that massive Pop.III stars would end up their lives with
spectacular explosions, many of which may have resulted in a GRB
(e.g., Fryer et al.~2001, Heger et al.~2002, 2003).  Their explosions likely
produced $\gamma$-ray bursts (GRBs), whose luminous afterglows were the
brightest, if short-lived, sources in the universe at that time.  Detection
and studies of this putative population of primordial GRBs opens some exciting
new prospects for cosmology.  
Cosmological uses of GRBs, including their potential as probes of the
reionization era, have been reviewed, e.g., by Lamb \& Reichart (2001),
Loeb (2002a,b), Djorgovski et al.~(2003b, 2004a), Hurley et al.~2004, etc.

Progenitors in the mass range $M_\star \sim 25 - 140 M_\odot$ and
$M_\star > 260 M_\odot$ would produce stellar mass black holes (BHs), and
presumably GRBs; the higher mass range is especially interesting, since higher
BH masses may imply higher accretion rates and thus higher GRB luminosities
(the so-called Type III collapsars).  In the progenitor mass range 
$M_\star \sim 140 - 260 M_\odot$, stars explode due to pair creation
instability, leaving no remnant (and thus without a GRB).

Thus, there is some real hope that significant numbers of GRBs and their
afterglows would be detectable in the redshift range $z \sim 6 - 20$, spanning
the era of the first star formation and cosmic reionization (Lamb \& Reichart
2001; Loeb 2002a,b; Bromm \& Loeb 2002; Djorgovski et al.~2003, 2004a).
This is supported by several studies in which photometric redshift indicators
for GRBs suggest that a substantial fraction (ranging from $\sim 10$\% to
$\sim 50$\%) of all bursts detectable by past, current, or soon forthcoming
missions may be originating at such high redshifts, even after folding in the
appropriate spacecraft/instrument selection functions (Fenimore \& Ramirez-Ruiz
2000; Reichart et al.~2001; Lloyd-Ronning, Fryer, \& Ramirez-Ruiz 2002; etc.).
We have every reason to hope that primordial GRBs from the reionization era
would be detectable with the SWIFT mission and the ground-based follow-up
studies of their afterglows.  The existing technology is well up to this task
(Lamb \& Reichart 2000, 2001; Ciardi \& Loeb 2000; Gou et al.~2004).

If GRBs reflect deaths of massive stars, their very existence and statistics
would provide a superb probe of the primordial massive star formation and the
IMF, as well as the transition from the Pop.III to Pop.II stars.
They would be by far the most luminous sources in existence at such redshifts
(much brighter than SNe, and most AGN), and they may exist at redshifts where
there were $no$ luminous AGN.  As such, they would provide unique new insights
into the physics, evolution, and early chemical enrichment of the primordial IGM
during the reionization era.

GRBs are more useful in this context than the QSOs, for several reasons.
First, they may exist at high redshifts where there were no comparably
luminous AGN yet.  Second, their spectra are highly predictable power-laws,
without complications caused by the broad Ly$\alpha$ lines of QSOs, and can
reliably be extrapolated blueward of the Ly$\alpha$ line.  Finally, they would
provide a genuine snapshot of the intervening ISM, without an appreciable
proximity effect which would inevitably complicate the interpretation of
any high-$z$ QSO spectrum: luminous QSOs excavate their Str\"omgren spheres
in the surrounding neutral ISM out to radii of at least a few Mpc, whereas
the primordial GRB hosts would have a negligible effect of that type.
Finally, they may hold a key for our understanding of the origin of SMBH
which power quasars.

\section{The Origins of Massive Black Holes}

We have discussed how there must be a deep connection between the formation
of galaxies and their central SMBHs, and outlined some of the evidence which
suggests that luminous QSOs at $z \sim 4 - 6$ must have started their growth
and chemical enrichment at consierably higher redshifts.  But where do the seed
BHs of these quasars -- and indeed the now ubiquitous (if quiescent) SMBHs in 
normal galaxies come from?

An excellent recent review of this subject is by Haiman \& Quataert (2004).
There are several possible physical paths towards the formation of seed BHs
which grow to power quasars.  First, there may be primordial BH remnants
from the Big Bang; this possibility sounds a bit contrived, and we direct a
curious reader to an excellent review by Carr (2003).  Another possibility is
a direct gravitational collapse of dark matter fluctuations, which is hard
to accomplish without some high-density seeds (non-Gaussian fluctuations),
and it would require that fragmentation into primordial stars is somehow
avoided; while this is in principle possible, there is no evidence or even an
independent need for such hypothetical scenario.  There is also a possibility
of a gravitational core collapse of relativistic star clusters which may have
been produced in early starbursts (see, e.g., Shapiro 2004, for a recent review
and references), which might lead to a formation of BHs in a mass range
$M_\bullet \sim 10^2 - 10^4 M_\odot$.

Probably the most secure astrophysical mechanism for the production of seed BHs
is as remnants of massive Pop.III star explosions, producing BHs in a mass range
$M_\bullet \sim 10^1 - 10^2 M_\odot$.
Thus, GRBs from Pop.III stars may also represent the signal events announcing
the birth of seed BHs, some or all of which eventually grow to the SMBH scales,
$M_\bullet \sim 10^6 - 10^9 M_\odot$.

BHs produced by Pop.III stars can accrete material, and act as ``mini-quasars''
(Madau \& Rees 2001; Madau et al.~2004).
Their UVX emission may be a significant contributor to the early reionization,
and indeed may be necessary if destructive feedback from massive Pop.III stars
and their explosions was too effective (Ricotti \& Ostriker 2004).

Once a seed BH is formed, it has to be grown, in some cases up to 
$M_\bullet \sim 10^9 M_\odot$ by $z \sim 6$.  This is not a trivial task,
given the limited length of time available, a few $\times 10^8$ yr
(see, e.g., Haehnelt \& Rees 1993).
Seed BHs can grow bigger in two ways: by accretion, and by merging.  Both
processes are possible, and the only question is which one dominates in
which range of masses, times, and environments.  Models of SMBH growth
within the standard hierarchical structure formation include, e.g.,
Haehnelt et al.~(1998), Bromley et al.~(2004), Volonteri et al.~(2003a,b),
Islam et al.~(2003), Haiman (2004), Haiman et al.~(2004), Yoo \& Miralda
Escud\'e (2004), etc.  BH mergers open a possibility
of detection of bursts of gravity waves, e.g., with the LISA mission
(http://lisa.jpl.nasa.gov).

A significant constraint on the models of SMBH growth is posed by the excellent
correlations between the SMBH masses and the global dynamical properties of
their host galaxies.  This indicates that the growth is driven by the processes
occurring on spatial scales up to nine orders of magnitude larger than the
Schwarzschild radius, rather than the physics in the immediate vicinity of
the BH.  While it is possible that some radiative or mechanical feedback
generated by the accretion machinery near the BH can affect the star formation
and gas infall and outflows at larger scales, it would have to be remarkably 
effective, and is also unlikely to affect the
dark halo masses -- and yet there is an excellent correlation between the
$M_\bullet$ and $M_{halo}$ (Ferrarese 2002a).

Whether the seed BHs come from the stellar/GRB remnants ($M_\bullet \sim 10 -
100 M_\odot$) or gravitational collapse of relativistic star clusters
($M_\bullet \sim 10^3 - 10^4 M_\odot$), an interesting question is whether
there are any ``intermediate'' mass ($M_\bullet \sim 10^3 - 10^5 M_\odot$)
still present today.  A good review of this subject is given, e.g., by
Miller \& Colbert (2004).  There are strong selection effects {\it against}
detection of such objects: their masses are too small to affect stellar dynamics
in galaxies  at a significant level, and their accretion luminosities -- if
any -- may be fairly low.  Two possibilities have been suggested in the
literature:
BHs in some globular clusters (but the analysis of data which suggested this
possibility has been challenged), and the so-called intermediate luminosity
X-ray sources seen in the vicinity of some nearby galaxies (at least some of
which turned out to be luminous background QSOs).  Thus, there is at this point
no conclusive evidence for the existence of intermediate mass BHs in the local
universe, but this is a search well worth pursuing.

\section{Concluding Comments.  The Cosmic Enlightment}

Cosmology is undergoing a remarkable period of growth and discovery.  Over the
past decade or two, we have witnessed a number of fundamental advances in our
understanding of the universe at large, and the origins and evolution of its
major baryonic constituents, galaxies and LSS.  Yet, it seems that we are now
entering an even more exciting period, exploration of the reionization era,
when the first luminous sources turned on and changed the universe.

In this short, and undoubtedly highly biased (just like the early structure
formation...) review, we outlined some ways in which the luminous manifestations
of BHs, from the GRBs to QSOs, can be used to illuminate crucial aspects of the
early star, galaxy, and LSS formation.  And while we seem to have at least some
understanding of the relevant astrophysical processes and the emerging big
picture of the reionization era, there are many outstanding problems and
challenges.

This field will likely keep the cosmologists gainfully(?) occupied for many
years to come.  Just like the synergy of the HST and ground-based 8 to 10-m
class telescopes changed the face of cosmology over the past dozen years,  
it is possible that the forthcoming space missions such as the JWST and the
new generation of extremely large (20 to 60-m class) ground-based optical/IR
telescopes (CELT/TMT, GMT, OWL), and large radio telescope arrays (EVLA, ALMA,
LOFAR, SKA, etc.) will lead to comparable advances in our understanding of the
birth of first stars, galaxies, and black holes.



\section*{Acknowledgments}
The author wishes to thank numerous collaborators, including
A. Mahabal, M. Bogosavljevic, M. Graham, R. Williams, D. Stern,
C. Baltay and the members of the Palomar-Quest Team, as well as
the grizzlied veterans of the DPOSS Team.
Many of the observational results described here were obtained at
Palomar and Keck Observatories, and we thank their staff for the
expert help during numerous observing runs.
This work was supported in part by the Ajax Foundation and grants
from the NSF and NASA.
This manuscript was completed while the author was enjoying the
hospitality and the stimulating atmosphere of the Institut
d'Astrophysique de Paris.
Finally, we thank the conference organizers for producing a 
wonderful meeting, and for the saintly patience while awaiting
this manuscript.


\end{document}